\documentclass[aps,10pt,showpacs,showkeys,twocolumn,preprint]{revtex4}
\usepackage{graphicx}
\usepackage{epsfig}
\usepackage{latexsym}
\usepackage{amsmath}

\begin{document}

\title{Noether symmetry in Horndeski Lagrangian}

\author{D. Momeni,\footnote{Corresponding author}}
  \email{momeni_d@enu.kz}
  \affiliation{Eurasian International Center for Theoretical Physics and Department of
General \& Theoretical Physics, Eurasian National University, Astana 010008, Kazakhstan}

\author{R. Myrzakulov}
  \affiliation{Eurasian International Center for Theoretical Physics and Department of
General \& Theoretical Physics, Eurasian National University, Astana 010008, Kazakhstan}

\date{\today}

\begin{abstract}
The Noether symmetry issue for Horndeski Lagrangian has been studied. We have been proven a series of theorems about the form of Noether conserved charge (current) for irregular (not quadratic) dynamical systems. Special attentions have been made on Horndeski Lagrangian. We have been proven that for Horndeski Lagrangian always is possible to find a way to make symmetrization.

\end{abstract}
arXiv:1410.1520 [gr-qc] 
\maketitle


\section{\label{b1}Introduction}
Canonical scalar fields are so popluar in theoretical physics because of their simplicity and easy way to interpret. Naturally if we use the Kaluza-Klein reduction for Einstein-Hilbert action in higher dimensions,  the reduced lower dimensional action is equal to a scalar theory which is coupled to an abelian gauge field. As an example, we can obtain Bergmann-Wagoner bi-scalar general action of scalar-tensor gravity \cite{Bamba:2014jua}. If we apply this reduction scheme on a more generalized model of gravity in the form of Lovelock gravity, we  obtain  more terms of scalar fields,which are now coupled to the gravity or to its second order 
invariants. Thechnically, as we know, when the brane model of Dvali-Gabadadze-Porrati (DGP) ~\cite{dgp} is decoupled, the resulted model is the scalar theory but with nonlinear terms of interaction~\cite{declim}. The idea of nonlinear scalar models are older than this new motivated idea. Indeed,Horndeski was who proposed the most general scalar field theory which its equation of motion (Euler-Lagrange (EL)) remains second order \cite{G. W. Horndeski} . As a fully covariant extension of the original Horndeski models, recently the idea of Galileon was introduced as the scalar theory with Galileon symmetry
\cite{Galileon}. This idea has been extended and developed through recent years to explain different aspects of gravitational theory from black hole physics to cosmology~\cite{gcos}-\cite{Naruko}. The models are written in a such way that they remain invariant under a local transformation of fields $\phi\to \phi+ \partial_{\mu}b$, here $b$ is gauge field.  A remarkable note about Galileon models is that they are represented the most generalized form of any other modified theory in the literature. The idea of Galileon is proposed in 
\cite{Galileon} and later it was extended to covariant form ~\cite{CovariantGalileon}. Other extensions have been followed  
~\cite{GeneralizedGalileon}.\par
In particular, the first two terms of Horndeski Lagrangian are very important to study. These terms  are  constructed from the second order forms like
 $(\nabla_{\mu}\phi)^2$ and 
$(\nabla_{\mu}\phi)^2\nabla_{\mu}\nabla^{\mu}\phi$. We would like to write these types of the Lagrangian densities in following forms ~\citep{KGB,G-inf}:
\begin{eqnarray}
{\cal L}_2&=&k(\phi, X),
\\
{\cal L}_3&=&-G(\phi, X)\nabla_{\mu}\nabla^{\mu}\phi,
\end{eqnarray}

Here  $k$ and $ G$  are arbitary functions of field $\phi$ and its kinetic part $X\equiv -\partial_\mu\phi \partial^\mu\phi/2$. Other higher order terms can be constructed using different geometric quantities like 
 $R$ (the Ricci tensor), $G_{\mu\nu}$( the Einstein tensor),and higher derivatives of field.
Furthermore, we know that
 $G=X$, we obtain 
covariant Galileons \cite{CovariantGalileon}. In this paper, we consider a  class of Horndeski Lagrangian, which is presented by the following action

\begin{eqnarray}\label{Horndeski Lagrangian}
S_{tot}=\int{\frac{R}{2}\sqrt{-g}d^4x}+\sum_{i=2}^{5}\int{d^4x\sqrt{-g}{\cal L}_i}, \label{S}
\end{eqnarray}
Where diffrent Lagrangian densities have been defined by the following:

\begin{eqnarray}
&&{\cal L}_2=G_2(\phi,X),\\&&
{\cal L}_3=G_3(\phi,X)\nabla_{\mu}\nabla^{\mu}
\phi\\&&
{\cal L}_4=G_{4,X}(\phi,X)\Big[\{\nabla_{\mu}\nabla^{\mu}
\phi\}^2\\&&\nonumber-\nabla_{\alpha}
\nabla_{\beta}\phi\nabla^{\alpha}\nabla^{\beta}
\phi\Big]+RG_{4}(\phi,X),\\&&
{\cal L}_5=G_{5,X}(\phi,X)\Big[
\{\nabla_{\mu}\nabla^{\mu}
\phi\}^3\\&&\nonumber-3\nabla_{\mu}\nabla^{\mu}\phi\nabla_{\alpha}\nabla_{\beta}\phi\nabla^{\alpha}\nabla^{\beta}
\phi\\&&\nonumber+2\nabla_{\alpha}\nabla_{\beta}\phi\nabla^{\alpha}
\nabla^{\rho}\phi\nabla
_{\rho}\nabla^{\beta}\phi\Big]\\&&\nonumber-6G_{\mu\nu}\nabla^{\mu}\nabla^{\nu}\phi G_{5}(\phi,X)
\end{eqnarray}

Here $G_{i,X}(\phi,X)\equiv\frac{\partial G_{i}(\phi,X)}{\partial X}$ ,$R$ is the Ricci tensor, $G_{\mu\nu}$ is the Einstein tensor, also  we set $\kappa^2=8\pi G=1,c=1$. Our aim here is to addressee symmetry issue for Horndeski Lagrangian given in (\ref{Horndeski Lagrangian}).  In literature a paper \cite{Alberto Nicolis} existed that specifically discusses constraints on a general scalar field if you enforce only the literal galilean symmetry. In our work we'll consider Horndeski models and not Galilleons. In particular, the model doesn't respect Galileon symmetry . The functions $G_i$ given in  (\ref{Horndeski Lagrangian}) are arbitrary functions of $\phi,X$.  Furthermore, because Horndeski Lagrangian is constructed in a covariant form, so it is manifestly Lorentz invarint.

We have been investigated all possible Noether symmetries of such models in the cosmological FLRW model. Our plan in this work is as the following:\\
In Sec. \ref{NS} we review the fundamental theory of Noether symmetry for regular dynamical systems. In Sec. \ref{second-action} we are considering Horndeski Lagrangian with Noether symmetries. In Sec. \ref{new} we have been proven a sequence of theorems about Noether symmetries for higher order derivatives models, including Horndeski Lagrangian. We conclude in Sec. \ref{conclusion}.
\section{\label{NS} Review of Noether symmetry}
Let us to consider a dynamical system with $N$ configurational cordinates $q_i$ is defined by the Lagrangian  $L\equiv L(q_i, \dot{q}_i;t),\ \ 1\leq  i\leq N$. The set of EL equations for this dynamical system is written as $\dot{p}_i-\frac{\partial L}{\partial q_{i}}=0,\ \ p_{i}\equiv\frac{\partial L}{\partial \dot{q}_i}$. We mention here that up or down the index has the same meaning since we are working in the flat space. What we called it as 
{\it Noether Symmetry Approach} is the existence of a vector, Noether vector $\vec{X}$ \cite{noether2},\cite{cap3},\cite{noether3},\cite{noether4},\cite{cap4}:

\begin{equation}\label{17}
 X=\Sigma_{i=1}^{N}\alpha^i(q)\frac{\partial}{\partial q^i}+
 \dot{\alpha}^i(q)\frac{\partial}{\partial\dot{q}^i}\,{,}
 \end{equation}
  and a set of non-singular functions $\alpha_{i}(q_j)$, in a such a way that 
 the Lie derivative of Lagrangian vanishes on all points of the manifold  
(the tangent space of
configurations $T{\cal Q}\equiv\{q_i, \dot{q}_i\}$):
 \begin{equation}\label{19}
 L_X{\cal L}=0\,
 \end{equation}
The mentioned condition can be written in the following expanded form:
 \begin{equation}\label{18}
 L_X{\cal L}=X{\cal L}=\Sigma_{i=1}^{N}\alpha^i(q)\frac{\partial {\cal L}}{\partial q^i}+
 \dot{\alpha}^i(q)\frac{\partial {\cal L}}{\partial\dot{q}^i}\,{.}
 \end{equation}
From the phase-space point of view, existence of $\vec{X}$ 
 implies that the total phase flux  enclosed in a region of space, is conserved along $X$. In fact, it is an easy task to show that  (by taking into account the EL equations):
 \begin{equation}\label{20}
 \frac{d}{dt}\frac{\partial {\cal L}}{\partial\dot{q}^i}-
 \frac{\partial {\cal L}}{\partial q^i}=0\,{,}\ \ 1\leq i \leq N.
 \end{equation}
Consequently, we have:
 \begin{equation}\label{21}
 \Sigma_{i=1}^{N}\frac{d}{dt}\left(\alpha^i\frac{\partial {\cal
 L}}{\partial\dot{q}^i}\right)=L_X{\cal L}\,{.}
 \end{equation}
If we can find $\alpha_{i}$  by vanishing the coefficents of all powers of $\dot{q}^i$, then we will show that there exist a \texttt{global} conserved charge as the following:
 \begin{equation}\label{22}
 \Sigma_0=\Sigma_{i=1}^{N}\alpha^ip_i
 \end{equation}
In other words, the existence of Noether symmetry implies that the {\it Lie derivative of the Lagrangian} on a given vector field ${\bf X} $ vanishes, i.e.
\begin{equation}
\pounds_{\bf X} L = 0. \label{noether}
\end{equation} 
It has been proven that Noether symmetry is a powerful tool to study cosmological models in different models \cite{cap3}-\cite{hann}. In our article we explore Noether symmetries (\ref{noether}) for the Horndeski Lagrangian, given by  (\ref{S}).

\section{Noether symmetry for third order Horndeski Lagrangian }\label{second-action}
To have a more comprehensive result, let us to consider the following Lagrangian which was proposed as minimal G-inflation \cite{Kobayashi:2011nu}:
\begin{eqnarray}
\mathcal{L}_3=k(\phi,X)-G(\phi,X)Y\label{model2}.
\end{eqnarray}
Where we denote by $Y=\nabla_{\mu}\nabla^{\mu}\phi$. There is no simple way to reduce this Lagrangian to a simpler quadratic form, because of the appearence of the highly nonlinear term $Y$. To resolve this problem, we propose a couple of Lagrange multipliers $\{\lambda,\mu\}$ in the following forms:
\begin{eqnarray}
&&L=3a\dot{a}^2+a^3\Big[k(\phi,X)-G(\phi,X)Y\Big]\\&&\nonumber-a^3\Big(\lambda(X-\frac{1}{2}\dot{\phi}^2)+\mu(Y-\nabla_{\mu}\nabla^{\mu}\phi)\Big).
\end{eqnarray}
By varying the Lagrangian $L$ w.r.t to the $\{X,Y\}$ we obtain $\lambda=k_{,X}-YG_{,X},\mu=-G$, so the reduced Lagrangian is written as the following:
\begin{eqnarray}
&&L(a,\phi,X,Y;\dot{a},\dot{\phi})=3a\dot{a}^2+ a^3\Big[k(\phi,X)\\&&\nonumber-G(\phi,X)Y\Big]-a^3(X-\frac{1}{2}\dot{\phi}^2)(k_{,X}-YG_{,X}).
\end{eqnarray}
The appropriate set of the coordinates for configuration space is $q^i=\{a,\phi,X,Y\}$. 
We define a vector field $\vec{X}=\alpha\frac{\partial}{\partial a}+\beta\frac{\partial}{\partial \phi}+\gamma\frac{\partial}{\partial X}+\theta\frac{\partial}{\partial Y}+\dot{\alpha}\frac{\partial}{\partial \dot{a}}+\dot{\beta}\frac{\partial}{\partial \dot{\phi}}$, here the functions $\alpha^i=\{\alpha,\beta,\gamma,\theta\}$ are defined on configuration space, so we have the following system of PDEs as a result of (\ref{noether}) for the above point-like Lagrangian:
\begin{eqnarray}
&&\frac{\partial G}{\partial a}=\frac{\partial G}{\partial Y}=0,\\&&
\frac{\partial k}{\partial a}=\frac{\partial k}{\partial Y}=0,\\&&
3\alpha a^{-1}(k-X(k_{,X}-YG_{,X})-YG)\\&&\nonumber+\beta(k_{,\phi}-X(k_{,X\phi}-YG_{,X\phi})-YG_{,\phi})\\&&\nonumber+\gamma(-X(k_{,XX}-YG_{,XX})+YG_{,X})\\&&\nonumber
-\theta(XG_{,X}+G)=0\\&&
\alpha+2a\alpha_{,a}=0\\&&
3\alpha a^{-1}(k_{,X}-YG_{,X})+\beta (k_{,X\phi}-YG_{,X\phi})\\&&\nonumber+\gamma(k_{,XX}-YG_{,XX})\\&&\nonumber-\theta G_{,X}+2\beta_{,\phi}(k_{,X}-YG_{,X})=0\\&&
6\alpha_{,\phi}+a^2\beta_{,a}(k_{,X}-YG_{,X})=0\\&&
\alpha_{,X}=\alpha_{,Y}=0\\&&
\beta_{,X}(k_{,X}-YG_{,X})=0\\&&
\beta_{,Y}(k_{,X}-YG_{,X})=0.
\end{eqnarray}
We know that $p_a=6a\dot{a},\ \ p_{\phi}=a^3\dot{\phi}(k_{,X}-YG_{,X})$ and the corressponding Noether charge is written as the following: 
\begin{eqnarray}
\Sigma=6\alpha a \dot{a}+\beta a^3\dot{\phi}(k_{,X}-YG_{,X})=\Sigma_{0}.
\end{eqnarray}
The system of PDEs has three major class of exact solutions.
\par
\textbf{Class A:} The system has the following exact solutions if we impose $k_{,X}-YG_{,X}=0,\ \ G_{,X}\neq0$:
\begin{eqnarray}
&&\alpha=\frac{\alpha_0}{\sqrt{a}},\ \ \beta=\frac{\alpha_0 Y}{a\sqrt{a}}\frac{\beta_0(ch(\phi)-3f(\phi))}{f'(\phi)},\\&&
\gamma=-\frac{\alpha_0ch(\phi)}{a\sqrt{a}G_{,X}},\\&&
\theta=0.
\end{eqnarray}
Where $\{h(\phi),f(\phi)\}$ are arbitrary functions of $ \phi$.
The associated conserved Noether charge is $\Sigma_{A}=6\alpha_0 \sqrt{a} \dot{a}$ from here we find $a(t)=\Big[a_{0}^{3/2}+\frac{\Sigma_{A}}{4\alpha_0}(t-t_0)\Big]^{2/3}$.
So we conclude here that:\par
\emph{The third order action of G-inflation  presented by (\ref{model2})
 has Noether symmetry vector field:}

\begin{eqnarray}
&&\vec{X}=\frac{\alpha_0}{\sqrt{a}}\frac{\partial}{\partial a}+Y\frac{\alpha_0}{a\sqrt{a}}\frac{\beta_0(ch(\phi)-3f(\phi))}{f'(\phi)}\frac{\partial}{\partial \phi}\\&&\nonumber-\frac{\alpha_0ch(\phi)}{a\sqrt{a}G_{,X}}\frac{\partial}{\partial X}+\frac{d}{dt}\Big[\frac{\alpha_0}{\sqrt{a}}\Big]\frac{\partial}{\partial \dot{a}}\\&&\nonumber+\frac{d}{dt}\Big[Y\frac{\alpha_0}{a\sqrt{a}}\frac{\beta_0(ch(\phi)-3f(\phi))}{f'(\phi)}\Big]\frac{\partial}{\partial \dot{\phi}}.
\end{eqnarray}

So, the action is in the following form:
\begin{eqnarray}
&&S=\int{\sqrt{-g}d^4x \Big(\frac{R}{2}+f(\phi)\nabla_{\mu}\nabla^{\mu}\phi\Big)}\\&&\nonumber=\int{\sqrt{-g}d^4x\Big(\frac{R}{2}+2Xf'(\phi)\Big)}
\end{eqnarray}
{\it  It is important to mention here that the above Noether symmetrized model is written in the following equivalent form:}
\begin{eqnarray}
&&S=\int{\sqrt{-g}d^4x \Big(\frac{R}{2}-\frac{1}{2}\nabla_{\mu}\psi\nabla^{\mu}\psi\Big)}
\end{eqnarray}
Where $\psi=\pm\int{\sqrt{f'(\phi)}d\phi}$.
Equation of motion of a scalar field is obtained:
\begin{eqnarray}
&&\ddot{\Psi}+3H\dot{\Psi}=0
\end{eqnarray}
Which can be solved by $\psi=\frac{\psi_0}{\Big[a_{0}^{3/2}+\frac{\Sigma_{A}}{4\alpha_0}(t-t_0)\Big]^{2}}$. So,if we apply Noether symmetry method to the  $\mathcal{L}_3$ , the set of the EOMs is completely integrable.\par

\par 

\textbf{Class B:} We suppose that $k_{,X}=0,\ \ G_{,X}=0$. Consequently we have
\begin{eqnarray}
\alpha=\alpha(a,\phi),\ \ \beta=\beta(a,\phi),G=G(\phi),k=k(\phi).
\end{eqnarray}
Exact solution for PDEs are given by:
\begin{eqnarray}
&&\alpha=\alpha(a)=\frac{\alpha_0}{\sqrt{a}}\\&&
\beta=\beta(a)=\beta_0\frac{\alpha(a)}{a}\\&&
\theta=\theta_0g(\phi)Y\frac{\alpha(a)}{a}.
\end{eqnarray}
Where
\begin{eqnarray}
&&k(\phi)=k_0e^{-3\phi/\beta_0},\\&&
G(\phi)=Ce^{-3\phi/\beta_0}+\frac{\theta_0}{\beta_0}e^{-3\phi/\beta_0}\int{d\phi g(\phi)e^{3\phi/\beta_0}}.
\end{eqnarray}
Noeher vector is:

\begin{eqnarray}
&&\vec{X}=\frac{\alpha_0}{\sqrt{a}}\frac{\partial}{\partial a}+\beta_0\frac{\alpha(a)}{a}\frac{\partial}{\partial \phi}+\theta_0g(\phi)Y\frac{\alpha(a)}{a}\frac{\partial}{\partial Y}\\&&\nonumber+\frac{d}{dt}\Big[\frac{\alpha_0}{\sqrt{a}}\Big]\frac{\partial}{\partial \dot{a}}+\frac{d}{dt}\Big[\beta_0\frac{\alpha(a)}{a}\Big]\frac{\partial}{\partial \dot{\phi}}
\end{eqnarray}

So, the following third order Galileon Lagrangian has Noether symmetry:
\begin{eqnarray}
\mathcal{L}=k(\phi)-G(\phi)\nabla_{\mu}\nabla^{\mu}\phi.
\end{eqnarray}
or equivalently:
\begin{eqnarray}
S=-\frac{1}{2}\int{d^4x \sqrt{-g}\Big(XG_{,\phi}-\frac{1}{2}k(\phi)\Big)}
\end{eqnarray}
{\it  This form is the k-inflation model in the standard canonical form .}\citep{kinflation}\par
\par
\textbf{Case C}: There is another interesting analytic class of solutions when we put $\gamma=\theta=0,k_X-YG_X=\Psi_{,X}$. In this case,
we have the following solutions:

\begin{itemize}

   \item $\alpha=\beta=0,\Psi=F_1(a,\phi)$
In this family , the scalar field has no dynamics, because the kinetic term is absent. Consequently the solutions have no physical application as cosmological models and we can ignore them.

\item  $\alpha=0,\beta=F_2(a,\phi),\Psi=F_{3}(a)$
This class  is potentially very interesting case, because the physical action is reduced to the 
Einstein-Hilbert action in the presence of a type of fluid in which the pressure is given by  $p\equiv \Psi$. It is easy to show that all the EOMs are integrable with this type of Noether symmetry.

\item  $\alpha=\beta=0,\Psi=\Psi(a,\phi,X)$
This solution corresponds to a generalized k-inflationary model in which the scalar field is non-minimally coupled to the cosmological background via the arbitrary function $=\Psi(a,\phi,X)$.

\item  $\alpha=0,\beta=c_1,\Psi=F_4(a,X)$
The model is a purely kinetic k-inflation which is  non-minimally coupled to the background. The model is well studied in literature \cite{Scherrer:2004au}. Such models are deserved to be considered as a unified model for dark matter and dark energy. The model is considered as a valid alternative to the generalized Chaplygin gas models. Such models are considered as  dark energy component with a very slow varying sound speed and are compatible with the cosmic microwave background fluctuations on large angular scales. Furthermore, recently the authors showed that \cite{Barausse:2015wia}, such models areable to explain the sensitivity parameters for emission by compact-star binaries.

\item   $\alpha=0,\beta=F_5(\phi),\Psi=X\frac{F_{6}(a)}{F_{7}(a)^2}+F_{8}(a)$
We have a type of fluid in the cosmological background. It ihas been shown that the model is completely integrable for a set of the EOMs. It has been shown that \citep{Gomes:2015dhl}, the model has scaling solutions when Horndeski's field is coupled to the background. Scaling solutions are those that $\frac{\rho_{\phi}}{\rho_m}$ is constant. So, we can find a solution of the following equivalent equation $\frac{d\rho_{\phi}}{d\ln a}=\frac{d\rho_m}{d\ln a}$.

\end{itemize}

\section{Noether symmetry for higher derivative Horndeski Lagrangian}\label{new}
We saw in the previous sectionsthat all types of the modified gravity theories always have nonlinearity, due to the higher order derivatives of the fields. Historically Ostrogradski \cite{ostrogradeski} was the first who studied canonical formalism (Hamiltonian formalism) for a class of models with higher order derivatives (see \cite{prd1990} for a comprehensive review). In the case of Galileon, even if we work at the level of minimal models, with $\mathcal{L}_3$, there is no simple way to reduce point-like Lagrangian to the standard quadratic form $\mathcal{L}=\mathcal{L}(\phi,X^m)$. A way is to introduce a set of appropriate Lagrange multipliers. But in this section we introduce an alternative to work with higher derivative Lagrangians. The simplest case which we are particularly interesting , is a class of Lagrangian functions  $L(q_i,\dot{q}_{i},(\partial_{t})^{n}[q_{i}]),\ \ n\leq2$. It is possible to extend it to $n\leq3$, but such cases doesn't have simple physical interpretations. 
We would like to see how we can find generalized Noether symmetry for a Lagrangian with second order time derivatives , namely $L=L(q_i,\dot{q}_i,\ddot{q}_i),\ \ 1\leq i\leq N$. We are interesting in the cases in which by integration part-by-part one can not reduce the Lagrangian to the standard quadratic form $L'=L'(q_i,\dot{q_i})$. The first simple example is point-like Lagrangian of standard GR, that it contains $\ddot{a}$ due to the curvature term $R=\pm6(\frac{\ddot{a}}{a}+\frac{\dot{a}^2}{a^2})$ but we can omit second order derivative term $\ddot{a}$ using an integration part-by part. For G-inflation models if we pass to the higher terms, we need a generalized Noether symmetry. This is one of the most important motivation for us in this work.\par
Let us  start to  define a set of  appropriate conjugate momenta:
\begin{eqnarray}
&&p_i=\frac{\partial L}{\partial \dot{q}^i},\ \ r_i=\frac{\partial L}{\partial \ddot{q}^i}.
\end{eqnarray}
The generalized EL equation is given by:
\begin{eqnarray}
&&\frac{\partial L}{\partial q^i}-\frac{d}{dt}\Big(\frac{\partial L}{\partial \dot{q}^i}\Big)+\frac{d^2}{dt^2}\Big(\frac{\partial L}{\partial \ddot{q}^i}\Big)=0.
\end{eqnarray}
Or equivalently we write down it in the following form:
\begin{eqnarray}
&&\frac{\partial L}{\partial q^i}-\dot{p}_{i}+\ddot{r}_i=0\label{EL2}.
\end{eqnarray}
We define a vector field:

\begin{eqnarray}
&&\vec{X}=\Sigma_{i=1}^{N}\Big(\alpha_i\frac{\partial}{\partial q^i}+\dot{\alpha}_i\Big[\frac{\partial}{\partial \dot{q}^i}-2\frac{d}{dt}\Big(\frac{\partial}{\partial \ddot{q}^i}\Big)\Big]\\&&\nonumber-\ddot{\alpha}_i\frac{\partial L}{\partial \ddot{q}^i}\Big)\label{v2}.
\end{eqnarray}

We call it as generalized Noether symmetry if and only if it satisfies the following algebrically vector equation:
\begin{equation}
 L_X{\cal L}=0\,,
 \end{equation}
In this case we find that the following polynomial should be vanish:

\begin{eqnarray}
&&\Sigma_{i=1}^{N}\alpha_i\frac{\partial L}{\partial q_i}+\Sigma_{i=1}^{N}\Sigma_{j=1}^{N}(\alpha_i)_{,j}\dot{q}_j\Big[\frac{\partial L}{\partial \dot{q}_i}\\&&\nonumber-2\frac{d}{dt}\Big(\frac{\partial L}{\partial \ddot{q}_i}\Big)\Big]-\Sigma_{i=1}^{N}\Big(\Sigma_{j=1}^{N}(\alpha_i)_{,j}\ddot{q}_j\\&&\nonumber+\Sigma_{j,k=1}^{N}(\alpha_i)_{,j,k}\dot{q}_j\dot{q}_k\Big)\frac{\partial L}{\partial \ddot{q}_i}=0.
\end{eqnarray}

If we collect all terms of different powers of $\{\dot{q}_i,\ddot{q}_j\}$ we obtain a system of second order PDEs (not first order like the standard Lagrangians) for $\{a_{i}(q^k)\}.$ Consequently, the associated Noether charge is obtained by the following theorem:\par
\texttt{Theorem}: {\it  For Lagrangian $L=L(q_i,\dot{q}_i,\ddot{q}_i)$, there exists a Noether vector symmetry  given by (\ref{v2}) 
and a Noether conserved charge: }
\begin{eqnarray}
&&K=\Sigma_{i=1}^{N}(\alpha_i p_i-\partial_t(\alpha_i r_i)).
\end{eqnarray}
\texttt{Proof}:
Using (\ref{EL2}) it is easy to  show that $\dot{K}=0$, because we've:

\begin{eqnarray}
&&\dot{K}=\Sigma_{i=1}^{N}(\partial_t(\alpha_i p_i)-\partial_{tt}(\alpha_i r_i)) \\&&\nonumber=\Sigma_{i=1}^{N}(\alpha_i\dot{p}_i+\dot{\alpha}_i p_i\-(\alpha_i \ddot{r}_{i}+2\dot{\alpha}_i\dot{r}_i+\ddot{\alpha}_ir_i))\\&&\nonumber
	\Longrightarrow \Sigma_{i=1}^{N}( \alpha_i(\dot{p}_i-\ddot{r}_i)+\dot{\alpha}_i(p_i-2\dot{r}_i)
-\ddot{\alpha}_ir_i)\\&&\nonumber=\Sigma_{i=1}^{N}\Big(\alpha_i\frac{\partial L}{\partial q_i}+\dot{\alpha}_i\Big[\frac{\partial L}{\partial \dot{q}_i}-2\frac{d}{dt}\Big(\frac{\partial L}{\partial \ddot{q}_i}\Big)\Big]-\ddot{\alpha}_i\frac{\partial L}{\partial \ddot{q}_i}\Big)\\&&\nonumber	\Longrightarrow  L_X{\cal L}=0.
\end{eqnarray}

This is our Q.E.D. \par
It is adequate to present the following generalized theorem for the dynaminal system in the form $L=L (q_i, q_ {I} ^ {(a)}; t), \ \ q_{i}^{(a)}=(\partial t) ^a [q_ {I}], \ \ 1\leq i\leq N, \ \ 1\leq a\leq s, \ \ s\leq N$. We are remembering to the mind that in this class of models, the generalized EL equation is written as the following:
\begin{eqnarray}
\frac{\partial L}{\partial q_{i}}+\Sigma_{a=1}^{s}(-1)^a(\partial t)^a(p^{a}_i)=0,
\end{eqnarray}
{ \it Here $p^{a}_i\equiv \frac{\partial L}{\partial q_{i}^{(a)}}$ is the new set of conjugate momenta}.\\ \\
\texttt{Theorem}: {\it  For Lagrangian $L=L (q_i, q_ {I} ^ {(a)}; t) $, there exists a generalized conserved Noether current given by:
\begin{eqnarray}
\mathcal{K}=\Sigma_{i=1}^{N}\Sigma_{a=1}^{s}(-1)^{a+1}(\partial t)^{a-1}\Big[\alpha_i p^{a}_i\Big].
\end{eqnarray}
\par
\texttt{Proof}: Using the Leibniz rule for derivatives we obtain:

\begin{eqnarray}
&&\partial_{t}(\mathcal{K})=\Sigma_{i=1}^{N}\Sigma_{a=1}^{s}(-1)^{a+1}(\partial t)^{a}\Big[\alpha_i p^{a}_i\Big]	\\&&\nonumber=\Sigma_{i=1}^{N}\Sigma_{a=1}^{s}\Sigma_{k=1}^{a}(-1)^{a+1}\frac{a!}{k!(a-k)!}\\&&\nonumber(\partial t)^k[\alpha_i](\partial t)^{a-k}[p^{a}_i]\\&&\nonumber	=\Sigma_{i=1}^{N}\Sigma_{a=1}^{s}\Sigma_{k=1}^{a}\Big((-1)^{a+1}\frac{a!}{k!(a-k)!}(\partial t)^k\\&&\nonumber\times[\alpha_i](\partial t)^{a-k}[\frac{\partial }{\partial q_{i}^{(a)}}]\Big) L	\Longrightarrow L_{\vec{X}}L=0
\end{eqnarray}

Where the following vector field,

 \begin{eqnarray}
&&\vec{X}=\Sigma_{i=1}^{N}\Sigma_{a=1}^{s}\Sigma_{k=1}^{a}\Big((-1)^{a+1}\frac{a!}{k!(a-k)!}(\partial t)^k\\&&\nonumber[\alpha_i](\partial t)^{a-k}\times[\frac{\partial }{\partial q_{i}^{(a)}}]\Big)
\end{eqnarray}

 is the Noether vector symmetry.\par
\texttt{Coroally}: {\it For a general Horndeski model with the following Lagrangian}
\begin{eqnarray}
L=L(\phi,(\nabla)^a\phi), 1\leq a<s.
\end{eqnarray}
The following vector is conserved:
\begin{eqnarray}
\mathcal{K}^{\mu}=\Sigma_{a=1}^{s}(-1)^{a+1}(\nabla)^{a-1}\Big[\alpha(\phi) \frac{\partial L}{\partial [(\nabla_{\mu})^a\phi]}\Big].
\end{eqnarray}
i.e. $\nabla_{\mu}\mathcal{K}^{\mu}=0$. In the above theorem if we use the conventional Horndeski's notations we should identify , $\nabla\equiv \nabla_{\mu},\ \ X\sim (\nabla_{\mu}\phi)^2,\ \ Y\sim (\nabla_{\mu})^2\phi=\nabla_{\mu}\nabla^{\mu}\phi$ and etc.\\ \\
\texttt{Illustrative example}: If we consider the minimal model of Horndeski theory, $\mathcal{L}_2+\mathcal{L}_3=k(\phi,X)-G(\phi,X)\nabla_{\mu}\nabla^{\mu}\phi$,   we obtain:

\begin{eqnarray}
&&\mathcal{K}^{\mu}=\nabla^{\mu}\Big[\alpha(\phi)G(\phi,X)\Big]
-\alpha(\phi)(k_{,X}\\&&\nonumber-G_{,X}\nabla_{\mu}\nabla^{\mu}\phi)\nabla^{\mu}\phi.
\end{eqnarray}

Which is trivially conserved if we fix $\alpha$ by Noether conservation vector condition (\ref{noether}).}

\section{Conclusions} \label{conclusion}
The most general form of scalar-tensor theory for gravity in the covariant form was proposed by Horndeski. The significant feature is that the set of equations of motion remains second order. Because Horndeski Lagrangian was constructed in a covariant form, consequently it is manifestly Lorentz invariant. In this work we addressed Noether symmetry of point like Lagrangian in the framework of Horndeski theory. We proposed a theorem about the Noether symmetry for a general highr order Lagrangian, specially in the form of Horndeski models. 
We extended the idea of the Lie generator of the normal tangent space. We have been proven that a vector field :

\begin{eqnarray} &&\vec{X}=\Sigma_{i=1}^{N}\Sigma_{a=1}^{s}
\\&&\nonumber\Sigma_{k=1}^{a}\Big((-1)^{a+1}\frac{a!}{k!(a-k)!}(\partial t)^k[\alpha_i](\partial t)^{a-k}[\frac{\partial }{\partial q_{i}^{(a)}}]\Big) 
\end{eqnarray} 

is the Noether vector symmetry for $L=L (q_i, q_ {I} ^ {(a)}; t) $. For a general Galileon model $\mathcal{L}_2+\mathcal{L}_3$,we have been proven that there exists conserved current. Our work explore more features of such extended models
\section{Acknowledgment} Thanks to {\bf Gregory Horndeski } for his great inspirations.

\end{document}